\documentclass[aps,prl,reprint,showpacs,superscriptaddress,groupedaddress]{revtex4-1}  
\usepackage{graphicx}  
\usepackage{dcolumn}   
\usepackage{bm}        
\usepackage{amssymb}   
\begin{document}


\title{Tunable resistivity of individual magnetic domain walls}

\author{J.~H.~Franken}
\author{M.~Hoeijmakers}
\author{H.~J.~M.~Swagten}
\author{B.~Koopmans}
\affiliation{Department of Applied Physics, Center for NanoMaterials and COBRA Research Institute, Eindhoven University of Technology, P.O.~Box~513,
5600 MB Eindhoven, The Netherlands}

\date{\today}

\begin{abstract}
Despite the relevance of current-induced magnetic domain wall (DW) motion for new spintronics applications, the exact details of the current-domain wall interaction are not yet understood. A property intimately related to this interaction is the intrinsic DW resistivity. Here, we investigate experimentally how the resistivity inside a DW depends on the wall width $\Delta$, which is tuned using focused ion beam irradiation of Pt/Co/Pt strips. We observe the nucleation of individual DWs with Kerr microscopy, and measure resistance changes in real-time. A $1/\Delta^2$ dependence of DW resistivity is found, compatible with Levy-Zhang theory. Also quantitative agreement with theory is found by taking full account of the current flowing through each individual layer inside the multilayer stack.
\end{abstract}

\maketitle

Current-induced motion of domain walls (DWs) in magnetic nanowires is an actively investigated topic \cite{Boulle2011}, because of possible application in memory and logic devices \cite{Parkin2008}. Although successful DW motion has been shown, details of the interaction between current and DWs remain unclear. In particular, the magnitude of the non-adiabatic torque varies greatly, especially in the case of perpendicularly magnetized materials \cite{Koyama2011,Boulle2008a,Burrowes2009}, and also theoretical work \cite{Zhang2004,Tatara2004,Vanhaverbeke2007} is focusing on various underlying physical mechanisms. It is suggested that this non-adiabatic contribution could arise when there is mistracking between the spin of conduction electrons and the local magnetization, which is also believed to be the origin of the electrical resistance induced in DWs. Systematic measurements of DW resistivity, which is the focus of the present paper, are therefore of crucial importance in unraveling the physics behind non-adiabatic spin torque.

Several effects can lead to resistance changes when a domain wall is introduced into a magnetic system. One effect is the anisotropic magnetoresistance (AMR), which always occurs when the component of magnetization parallel to the current flow changes and is therefore not intrinsic to the DW. Various mechanisms can lead to \emph{intrinsic} DW resistivity (DWR) \cite{Marrows2005}. Viret et al. \cite{Viret1996} first treated resistance due to spin mistracking semi-classically; the DW represents a magnetic field rotating over a distance $\Delta$ (DW width) in the rest frame of an electron moving at the Fermi velocity $v_{\rm{F}}$, and the electron spin can only follow this field adiabatically if the precession frequency ($2J_{\rm{sd}}/\hbar$) of the spin about the exchange field is much larger than the rotation frequency ($\pi v_{\rm{F}} / \Delta$). Levy and Zhang later presented a quantum-mechanical version of this model \cite{Levy1997}, starting from the same Hamiltonian used to describe the giant magnetoresistance (GMR) in magnetic multilayers. Mistracking leads to mixing of the majority and minority spin channels, changing the spin dependent scattering at impurities within the DW. This increases the resistance of one spin channel while reducing that of the other, thereby giving a higher net resistance of the two parallel channels. Some experimental values of DWR compatible with the theory of Levy and Zhang have been reported \cite{Ravelosona1999,Viret2000,Marrows2004,Aziz2006}, although measurements were often hampered by the presence of other magnetoresistive effects. In contrast, other conflicting theories of DWR predict negative resistance \cite{Tatara1997} or a contribution that can have either sign \cite{VanGorkom1999}. Unusually high DWR values were found in epitaxial Co wires at 77\,K \cite{Ebels2000}, which was attributed to the spin flip length being much larger than the width of the DW leading to spin accumulation at the DW, although it was later argued that this is not sufficient to explain their results \cite{Simanek2001}. Despite this considerable progress during the past years, the width of the DW, which is obviously a crucial parameter in fundamentally unraveling the origin of DWR, could never be changed systematically in experiments on DW resistivity.

In this Letter we address this outstanding issue, by introducing a unique approach to simultaneously engineer the DW width and measure its intrinsic resistivity. We use Ga irradiation to create a controlled magnetic domain pattern in a Pt/Co/Pt strip \cite{Aziz2006}, and measure the resistance of individually appearing DWs directly by real-time magneto-optic imaging. Interestingly, the perpendicular anisotropy is tuned by the irradiation dose, and thereby the DW width $\Delta$ can be carefully controlled. It is found that the DWR scales as $1/\Delta^2$, lending strong support to the aforementioned Levy-Zhang theory \cite{Levy1997}. Also quantitative agreement is found by taking into account the current flow through the individual layers of the multilayer stack \cite{Cormier2010}, allowing us to estimate the spin asymmetry of the current in the Co layer. We believe that the observed dominant role of mistracking between mobile electron spins and local DW magnetic moments will aid in pinpointing the origin of non-adiabatic spin-transfer-torque in novel DW devices.

\begin{figure}
\includegraphics[width=\linewidth]{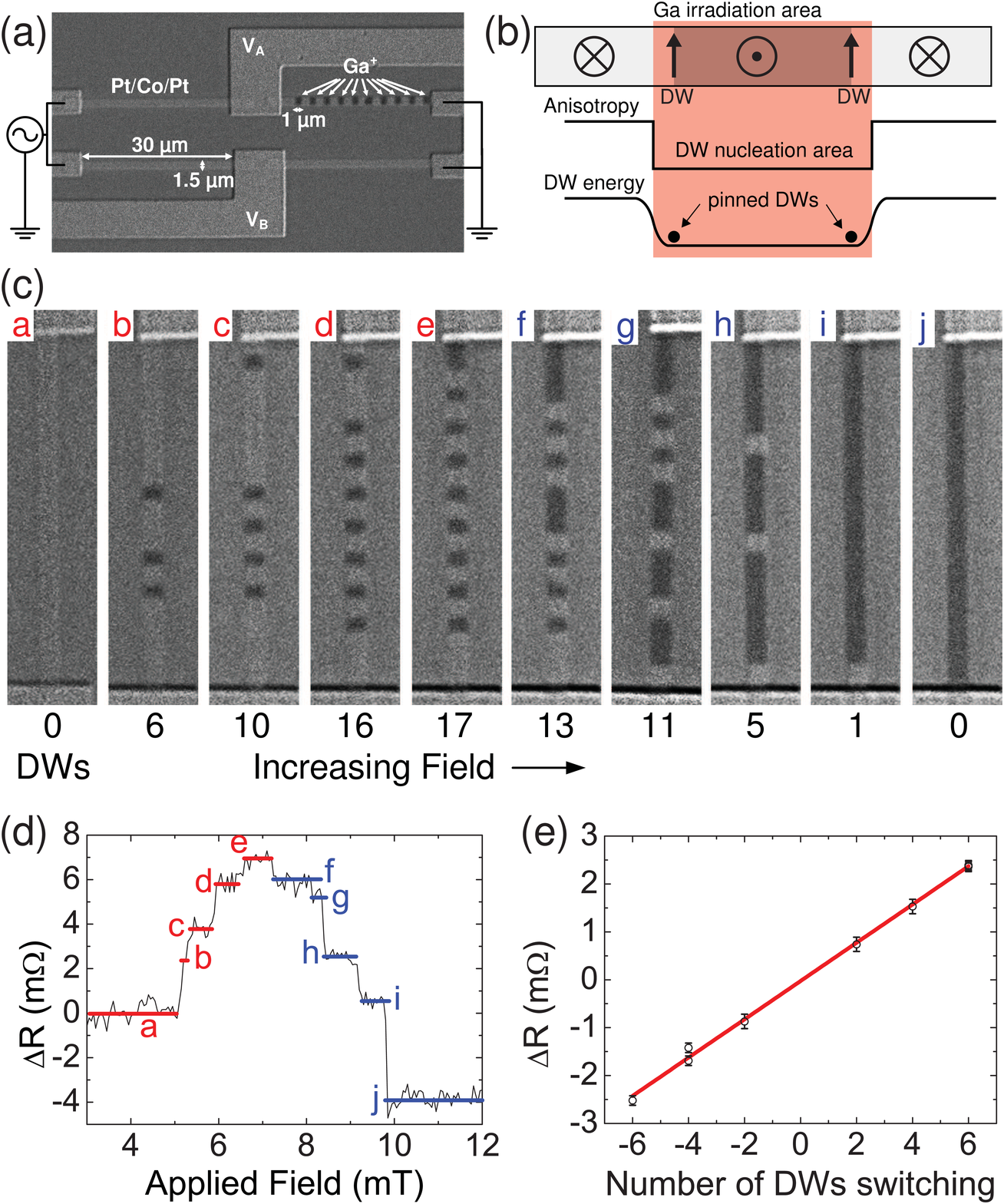}
\caption{\label{Figure1} (a) Kerr microscopy image of the experimental geometry showing 4 Pt/Co/Pt strips in a Wheatstone bridge configuration, where one of the strips has been patterned using a focused Ga-ion beam with an irradiation dose of $0.56\times 10^{13}$\,Ga ions/cm$^2$. The dark areas have inverted magnetization and correspond to the irradiated areas. (b) Close-up of a Ga irradiated area (top), leading to an anisotropy well along the wire direction (middle), which translates into an energy well for the two DWs that are nucleated in the irradiated area. The DWs will thus remain within the irradiated area where their width can be tuned by the irradiation dose. (c) Kerr images recorded upon increasing the external magnetic field, showing nucleation and annihilation of magnetic domains. (d) Resistance as a function of field. Discrete steps are observed whenever DWs are nucleated or annihilated as visualized in (c). (e) Magnitude of resistance jumps as a function of number of DWs nucleating/annihilating. The slope represents the resistance increase due to a single DW. }
\end{figure}

The resistance of DWs is measured on 1.5\,$\mu$m wide Pt (4\,nm) / Co (0.5\,nm) / Pt (2\,nm) strips fabricated by e-beam lithography, sputter deposition and lift-off. We adopt an on-sample Wheatstone bridge configuration \cite{Aziz2006} with 4 identical Pt/Co/Pt strips  as shown in Fig.~1(a). One of the four strips is then patterned with 30 keV Ga ions using a focused ion beam, to create regions with a reduced magnetic anisotropy and coercivity \cite{Aziz2006,Franken2011a}. The resistance of this strip changes when DWs are present, leading to a change in the offset voltage $V_{\rm{A}} - V_{\rm{B}}$. An AC current runs in the indicated direction and $V_{\rm{A}} - V_{\rm{B}}$ is measured using a lock-in amplifier. Knowing the resistance of the four individual strips ($R \sim1.3\,$k$\Omega$), the resistance change in the patterned strip $\Delta R$ can be accurately determined.

Starting from negative perpendicular saturation, $\Delta R$ is measured as a function of positive applied field (Fig.~1(d)), while the magnetic configuration is imaged in real-time in a wide-field Kerr microscope (Fig.~1(c)) \cite{Evico}. As the magnetic field is increased, domains are nucleated in the Ga-irradiated regions and expand to the irradiation boundaries, where DWs get pinned, as sketched in Fig.~1(b) \cite{Franken2011,Lavrijsen2010,Franken2011a}. Since there is some spread in the nucleation fields of each irradiated region, not all domains are created at the same time. Each time a new domain appears, a step in the resistance is observed, corresponding to the resistance of the new pair of DWs. When the field is increased further, DWs depin from the irradiation boundaries and annihilate with neighboring DWs, which is accompanied by a stepwise decrease of the resistance.

This technique is very powerful to directly determine the resistance of individual (pairs of) DWs, which was before only attempted indirectly by comparing with MFM images recorded afterwards \cite{Danneau2002,Aziz2006}. Furthermore, our measurement scheme allows for exclusion of measurement artifacts caused by other magnetoresistive effects. For example, the AMR contribution to the resistance does not depend on the number of domain walls, since the magnetic orientation within the Bloch DWs is always perpendicular to the current flow. Another measurement artifact is readily observed in Fig.~1: the final resistance step from (i) to (j) in Fig.~1(d) appears to originate from the disappearance of a single DW in Fig.~1(c). However, this resistance change is too large to correspond to a single DW, and instead originates from switching of the magnetic area underneath the bottom contact, probably yielding a contribution from the anomalous Hall effect (AHE). A similar effect is observed in the switching from state (d) to (e), albeit less prominent. Therefore, switching events that include a contact are excluded from the data analysis. In Fig.~1(e), all observed resistance jumps are plotted as a function of the number of DWs (dis)appearing, showing an expected linear behavior, from which the resistance of an individual DW can be accurately extracted.

\begin{figure}
\includegraphics[width=\linewidth]{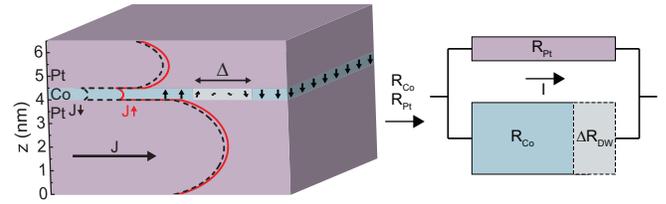}
\caption{\label{Figure2} Sketch of the multilayer geometry, including a calculation of the current distribution of majority and minority electrons, using the Fuchs-Sondheimer model. The same material parameters as Ref.~\cite{Cormier2010} were used. 3\% of the current flows through the Co layer, which was used to evaluate the resistance of the Pt and Co layers in a parallel resistor model. On the right, the contribution of the Co layer, Pt layers and DW resistance to the total resistance of the wire is sketched. }
\end{figure}

\begin{figure}
\includegraphics[width=\linewidth]{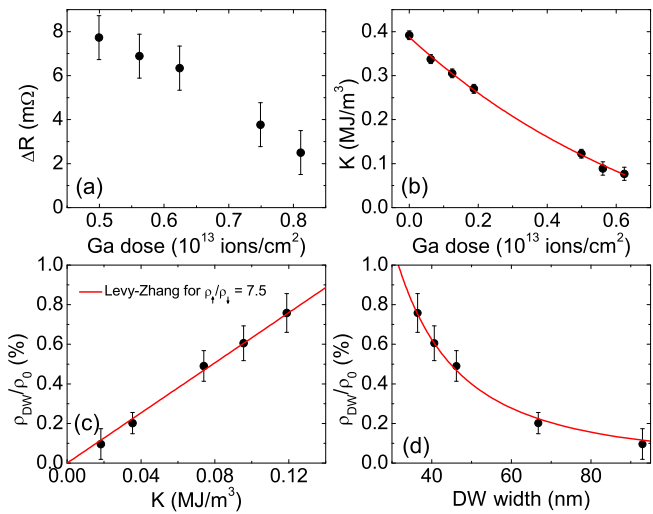}
\caption{\label{Figure3} (a) Resistance change $\Delta R$ due to 20 DWs in a Pt/Co/Pt strip as a function of Ga dose. (b) Perpendicular anisotropy as a function of Ga dose. The red line is an exponential fit. (c) Normalized DW resistivity as a function of anisotropy. The red line is the theoretical result of the Levy-Zhang model with $\rho_{\uparrow}/\rho_{\downarrow} = 7.5$. The same data is plotted in (d) as a function of DW width, showing the $1/\Delta^2$ dependence.}
\end{figure}

Combining all resistance steps linked to switching events, we find a positive resistance change $\Delta R = 7\pm1$\,m$\Omega$ when the maximum number of 20 DWs is present. To quantitatively compare these data with theoretical models we should obviously consider the DW \emph{resistivity} in the magnetic Co layer, excluding current shunting through the Pt layers. The spin-resolved current density in the layer system was therefore computed using a Fuchs-Sondheimer model with parameters from \cite{Cormier2010}, as shown in Fig.~2. We find a fraction $p \approx 3$\% of the current flows through the Co layer, much less than the 8\% predicted by only considering the bulk resistivities \cite{Aziz2006} without interface scattering. This fraction is used to compute $R_{\rm{Co}} = R/p$ and $R_{\rm{Pt}} = R/(1-p)$, which can then be used to quantify the resistance change of the Co layer $\Delta R_{\rm{DW}}$ due to all DWs present, using the geometry of Fig.~2,
    \begin{equation}
    \Delta R_{\rm{DW}} = \frac{\Delta R \,(R_{\rm{Pt}} +R_{\rm{Co}})}{R \,R_{\rm{Pt}} - R_{\rm{Co}} \Delta R} R_{\rm{Co}}.
    \end{equation}
Finally, the desired DW resistivity is then found by multiplying with the layer cross section $w \,t_{\rm{Co}}$ and dividing by the total width of the $N$ DWs,
    \begin{equation}
    \rho_{\rm{DW}} = \frac{\Delta R_{\rm{DW}}\, w\, t_{\rm{Co}}}{N\,\Delta}. \label{eq:roDW}
    \end{equation}
The DW width $\Delta$ is estimated by $\pi\sqrt{A/K}$, with $A=16\,$pJ/m \cite{Metaxas2007} the exchange stiffness and $K$ the magnetic anisotropy, whose magnitude we will discuss later.

As mentioned, the DW width $\Delta$ is a crucial parameter in theories on DWR. Therefore, we take a unique approach to tune the DW width by the Ga irradiation dose. It is known that Ga irradiation reduces the perpendicular magnetic anisotropy constant $K$ \cite{Warin2001,Franken2011a}, and this leads to wider DWs. Furthermore, we know that a Ga irradiation boundary acts as an energy barrier due to the sudden increase of magnetic anisotropy, and a pinned DW resides at the base of the barrier, hence in the irradiated region, as sketched in Fig.~\ref{Figure1}(b) \cite{Franken2011a,Franken2011,Lavrijsen2010}. Thus by varying the irradiation dose, we can vary the width of a pinned DW, of which the resistance is measured as explained before. Fig.~3 constitutes the main result of this experiment.

In Fig.~3(a), the raw data of the resistance increase $\Delta R$ due to 20 DWs in a Pt/Co/Pt strip is plotted as a function of Ga dose. A decrease is observed, meaning that a lower anisotropy yields a lower DW resistance. In Fig.~3(b), the anisotropy of Pt (4\,nm) / Co (0.5\,nm) / Pt (2\, nm) as a function of Ga dose is shown and fitted with exponential decay. The anisotropy was measured on irradiated 5$\,\mu$m wide Hall bars by fitting $M(H)$ according to Stoner-Wohlfarth theory, with $H$ applied at different angles to the film normal \cite{Franken2011a}. The result is used in Fig.~3(c) to convert each used Ga dose to a value of $K$ on the $x$-axis. On the $y$-axis, the DW resistivity $\rho_{\rm{DW}}$ (Eq.~\ref{eq:roDW}) normalized by the resistivity of the Co layer $\rho_0$ ($=R_{\rm{Co}} \times w \times t_{\rm{Co}} / L \approx 1.07\,\mu\Omega$m) is shown.

As one of the most prominent observations in this paper, it is seen that the DW resistivity as a function of anisotropy is not only remarkably linear, but also extrapolates through the origin. This implicitly means that the DW resistivity $\rho_{\rm{DW}}$ scales with $1/\Delta^2$, which we show in Fig.~3(d) by deducing the DW width from the anisotropy constants. This, together with the fact that $\rho_{\rm{DW}}>0$ for all DW widths, strongly suggests that the effect should arise from additional scattering due to mixing between the spin channels when electrons are trying to follow the changing magnetization direction within the wall, an effect that according to our data quadratically becomes larger when reducing the width of the wall. As already mentioned in the introduction, this is fully compatible with the spin-mistracking model proposed by Levy and Zhang \cite{Levy1997}, where
    \begin{eqnarray} \label{eq:LZ}
    	\frac{\rho_{\rm{DW}}}{ \rho_{0}} & = & \lefteqn{\frac{1}{5} \left(\frac{\pi \hbar^{2} k_{\rm{F}}} {4m\Delta J }\right)^2} \nonumber \\
    && \times \left(\frac{\rho_{\uparrow}}{ \rho_{\downarrow}} -2 + \frac{\rho_{\downarrow} }{ \rho_{\uparrow}} \right) \left(3+\frac{10 \sqrt{\rho_{\uparrow} / \rho_{\downarrow} }}{\rho_{\uparrow} / \rho_{\downarrow}+1}\right),
    \end{eqnarray}
with $\rho_{\uparrow(\downarrow)}$ the resistivity of up (down) electrons, $\hbar$ Planck's constant, $k_{\rm{F}}$ the Fermi wavevector, $m$ the electron mass, and $J$ the exchange splitting. Using $\Delta = \pi \sqrt{A/K}$, the linear dependence on $K$ as observed in Fig.~3(c) is recovered. Quantitative information can be gained from the slope of the linear fit, which yields a value for the spin asymmetry  $\rho_{\uparrow}/ \rho_{\downarrow} = 7.5$ if the other parameters are kept constant at $k_{\rm{F}}=1\,$\AA$^{-1}$, $A=16\,$pJ/m and $J=0.5$\,eV. We should note that this is the spin asymmetry of the current in the ultrathin Co layer, so a higher value in bulk Co is needed to reproduce this in the Fuchs-Sondheimer model. Still, the value  is very reasonable and comparable with previous measurements and thin-film band structure calculations \cite{Aziz2006}.

We strongly believe that the observed effect as a function of Ga dose is dominated by the tuning of the DW width. While Ga irradiation could influence the transport properties, we verified that the change in ordinary resistivity was only 0.5\% in the Ga dose range used. Furthermore, preliminary experiments on current-assisted DW depinning from the irradiated area indicate no decrease of spin torque efficiency with increasing dose, hence spin-polarized transport appears to be conserved. Finally, intrinsic magnetoresistance effects such as AMR or anisotropic interface magnetoresistance \cite{Kobs2011} can be excluded, since the resistance of a DW should in that case increase instead of decrease as a function of $\Delta$.

In conclusion, our measurement of the domain wall resistivity in Pt/Co/Pt as a function of magnetic anisotropy by variation of Ga irradiation dose lend strong support to the theory of Levy and Zhang \cite{Levy1997}. The $1/\Delta^2$ dependency predicted by the model was, to our knowledge, verified for the first time, and quantitative agreement is found with a value for the spin asymmetry $\rho_{\uparrow}/ \rho_{\downarrow} = 7.5$. Besides its fundamental importance, this could have interesting implications for current-induced domain wall motion, in particular the non-adiabatic spin-transfer torque (STT) contribution characterized by the $\beta$ parameter \cite{Zhang2004,Tatara2004,Vanhaverbeke2007}, in which both the width and the resistivity of the DW are important parameters. Our result implies that mistracking of the spin of conduction electrons with the local magnetization increases significantly in smaller DWs. Recent experiments have revealed that $\beta$ is relatively insensitive to the DW width down to 1 nm \cite{Burrowes2009}. This calls for further systematic studies of $\beta$ and the domain wall resistivity down to sub-nm DW widths, as a shift between the spin relaxation \cite{Zhang2004}, spin mistracking \cite{Vanhaverbeke2007} and momentum transfer \cite{Tatara2004} mechanisms could become apparent.

This work is part of the research programme of the Foundation for Fundamental Research on Matter (FOM), which is part of the Netherlands Organisation for Scientific Research (NWO).

\end{document}